\documentclass[conference]{IEEEtran}

\usepackage[numbers,sort&compress]{natbib}
\usepackage{graphicx}
\usepackage{orcidlink}
\usepackage{enumitem}
\usepackage{amsmath,amssymb,amsfonts}
\usepackage{algorithmic}
\usepackage{textcomp}
\usepackage{balance}
\usepackage[utf8]{inputenc}
\usepackage{url}
\usepackage[]{algorithm2e}
\usepackage{wrapfig}
\usepackage{caption}
\usepackage{subcaption}
\usepackage{hhline}
\usepackage{tabularx}
\usepackage{booktabs}
\usepackage{colortbl}
\usepackage{todonotes}
\usepackage[scaled=0.86]{helvet}
\usepackage{framed}
\usepackage{multirow}
\usepackage{rotating}
\definecolor{SkyBlue}{rgb}{0.53, 0.81, 0.92} 
\usepackage[framemethod=TikZ]{mdframed}
\usetikzlibrary{positioning,arrows.meta,shapes.geometric,fit,backgrounds}
\usepackage{minted}
\usepackage{tcolorbox}
\usepackage{pifont}
\usepackage{xcolor}

\definecolor{light-gray}{gray}{0.9}
\hypersetup{
    colorlinks=true,
    linkcolor=black,
    filecolor=magenta,      
    urlcolor=blue,
    citecolor=black,
    pdfborder={0 0 0}
}

\mdfdefinestyle{MyFrame}{%
    linecolor=black,
    outerlinewidth=0.15pt,
    roundcorner=3pt,
    innertopmargin=2pt,
    innerbottommargin=2pt,
    innerrightmargin=4pt,
    innerleftmargin=4pt,
    backgroundcolor=light-gray}
\mdfdefinestyle{MyFrameSimple}{%
    linecolor=black,
    outerlinewidth=0.15pt,
    roundcorner=3pt,
    innertopmargin=2pt,
    innerbottommargin=2pt,
    innerrightmargin=4pt,
    innerleftmargin=4pt}

\newcommand{\sectopic}[1]{\vspace*{0.1em}\par\noindent{\textit{\bfseries #1}}}
\newcommand{\StepOne}{{\color{red}\ding{202}}}
\newcommand{\StepTwo}{{\color{red}\ding{203}}}
\newcommand{\StepThree}{{\color{red}\ding{204}}}
\newcommand{\StepFour}{{\color{red}\ding{205}}}

\IEEEoverridecommandlockouts
\begin{document}

\title{EdgeFlow: Edge-Map Augmented VLM-Based Flowchart Processing for Industrial Requirements Engineering}

\author{
\IEEEauthorblockN{Zhifei Dou\,\orcidlink{0009-0006-2048-5895}}
\IEEEauthorblockA{Huawei Research Canada\\zhifei.dou@h-partners.com}
\and
\IEEEauthorblockN{Shabnam Hassani\,\orcidlink{0009-0008-3056-4073}}
\IEEEauthorblockA{Huawei Research Canada\\hassani.shabnam@h-partners.com}
\and
\IEEEauthorblockN{Ou Wei}
\IEEEauthorblockA{Huawei Research Canada\\ou.wei1@huawei.com}

}

\maketitle
\begin{abstract}
Flowcharts are widely used in industrial requirements, but usually remain embedded as static images. Vision Language Models (VLMs) show promise in the conversion of these flowcharts into machine-readable models for RE activities, yet, when directly applied to flowchart conversion,  they often fail on topology-critical visual details. To address this,  we propose \textsf{EdgeFlow} that augments a VLM's original input with a deterministically extracted Canny edge map--acting as a structural prior--to improve flowchart-to-Mermaid conversion, without requiring annotated training data or domain-specific model fine-tuning.

We evaluate \textsf{EdgeFlow} on \textit{IndusReqFlow}, a dataset sourced from real-world requirements. Compared with off-the-shelf VLMs, \textsf{EdgeFlow} improves node-level F1 by 17.39 percentage points and edge-level F1 by 16.94 percentage points. At the path level, \textsf{EdgeFlow} improves path F1 by 11.06 percentage points, enabling better support for model-based testing.   These results demonstrate that \textsf{EdgeFlow} provides a  practical, training-free means to improve topology-preserving flowchart-to-Mermaid conversion for industrial RE. Cross-dataset evaluation results on a public synthetic benchmark show no significant improvement; this highlights the need for diverse benchmarks incorporating  industrial data for the comprehensive evaluation of future VLM-based RE tools.

\end{abstract}

\section{\textbf{Introduction}}\label{sec:intro}

Flowcharts are widely used to communicate and validate system behavior, business logic, and operational procedures in specification documents~\cite{arbaz2024genflowchart,tannert2023flowchartqa,moody2009physics}.
Their primary advantage is practical: they provide an accessible, review-friendly abstraction of control flow (e.g., branching, exception handling, parallelism, termination). As a result, industry standards have advocated the use of flowcharts  to facilitate  requirements discussions across roles and domains during requirements analysis, architecture design, and test case generation~\cite{baltes2014sketches,iso5807}. 

Requirements Engineering (RE) community has long studied modeling notations for capturing behavior and processes, specifically to enable automated verification and validation (e.g., deriving tests from behavioral models)~\cite{briand2002umltesting,utting2010practicalmbt,elfar2002modelbased}. However, flowcharts frequently remain embedded as static images within documents. 
This creates a visual specification gap that limits the utility of these artifacts for RE automation~\cite{deka2025flowchart2mermaid,singh2024flowvqa,arbaz2024genflowchart,deka2026bpmn}.

To date, RE research has successfully leveraged textual processing,  ranging from NLP-based extraction~\cite{zhao2020nlp,arora2016domain} to LLM-driven model generation~\cite{debari2024uml,eisenreich2025llm,ferrari2024modelgen}; however, the extraction of structural models from visual artifacts remains a critical bottleneck. 
The lack of machine-interpretable representations for flowchart-embedded requirements hinders both core RE tasks---such as traceability checking and change impact analysis---and downstream automation, including model-based test generation and coverage analysis~\cite{briand2002umltesting,utting2010practicalmbt,elfar2002modelbased}. This creates a need for methods capable of converting visual specifications into path-enumerable structures while preserving control-flow semantics. While this visual specification gap affects various notation types, this paper focuses on flowcharts given their widespread application in industrial documentation.

Recent Vision Language Models (VLMs) have shown strong multimodal reasoning capabilities across software engineering artifacts; however, their application to flowchart parsing is hindered by a critical limitation: the misperception of topological structures, specifically node connectivity and edge direction~\cite{pan2024flowlearn,singh2024flowvqa,omasa2025arrow,deka2026bpmn}. This is particularly consequential for path-oriented analysis in system development process, such as test case generation.
While existing mitigation strategies attempt to decompose processing into staged pipelines using OCR or segmentation~\cite{arbaz2024genflowchart}, these approaches remain sensitive to upstream perception failures.

\vspace{-0.5ex}
\subsection{\textbf{Motivation}}
Through our collaboration with Huawei's production-line testing teams across different industrial domains, we observe that requirements documents frequently specify operational procedures and decision logic using flowcharts. 
For production-line validation, engineers routinely need to derive test cases that cover the feasible paths (e.g., normal flows, exception handling, and alternative branches) encoded in these flowcharts to understand system requirements and perform model-based testing (MBT)~\cite{briand2002umltesting,elfar2002modelbased,utting2010practicalmbt}. 
Currently, such path enumeration is performed manually---a labor-intensive and error-prone process that scales poorly for complex flowcharts and scenarios with frequent revisions.

To bridge this gap, we propose \textsf{EdgeFlow} (\textit{VLM- based Flowchart Processing with Edge-Map Augmentation}), a training-free framework that augments a VLM's input with deterministic edge maps. \textsf{EdgeFlow} uses Canny edge detection~\cite{canny1986computational} (Section~\ref{sec:background}) to supply the VLM with high-frequency topological signals, acting as \emph{structural priors} that bias inference toward geometric features~\cite{zhang2023adding}. 
Unlike staged pipelines that require external object detectors~\cite{omasa2025arrow} or segmentation models, \textsf{EdgeFlow}'s edge-map extraction is fully deterministic and requires no annotated training data, facilitating industrial deployment.

\vspace{-0.5ex}
\subsection{\textbf{Illustrative Example}}
\label{subsec:illustrative_example}
\begin{figure}[t]
\centering
  \begin{subfigure}{\linewidth}
    \centering
    \includegraphics[width=0.98\linewidth]{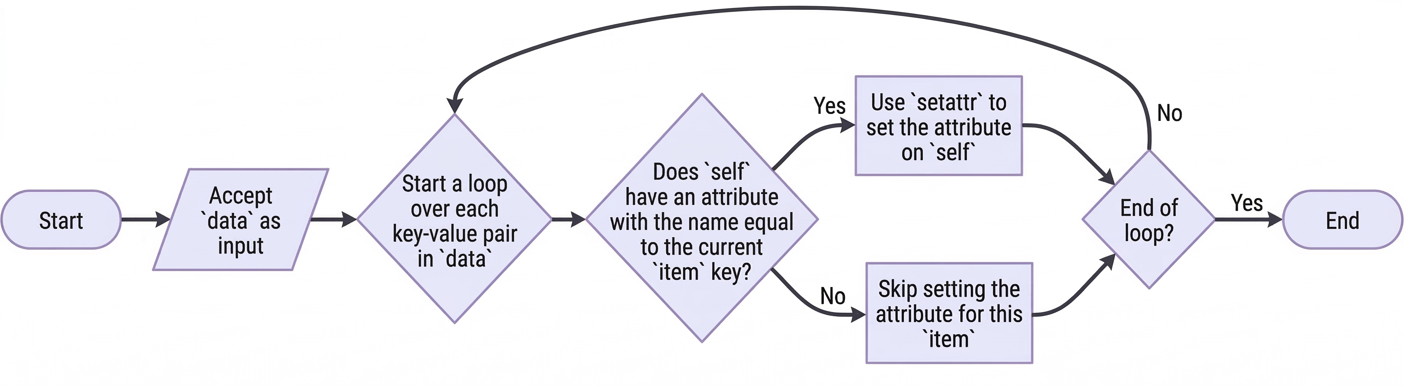}
    \caption{Input flowchart (FlowVQA example)}
    \label{fig:example_images_raw}
  \end{subfigure}\\[1.5ex]
  \begin{subfigure}{\linewidth}
    \centering
    \includegraphics[width=0.98\linewidth]{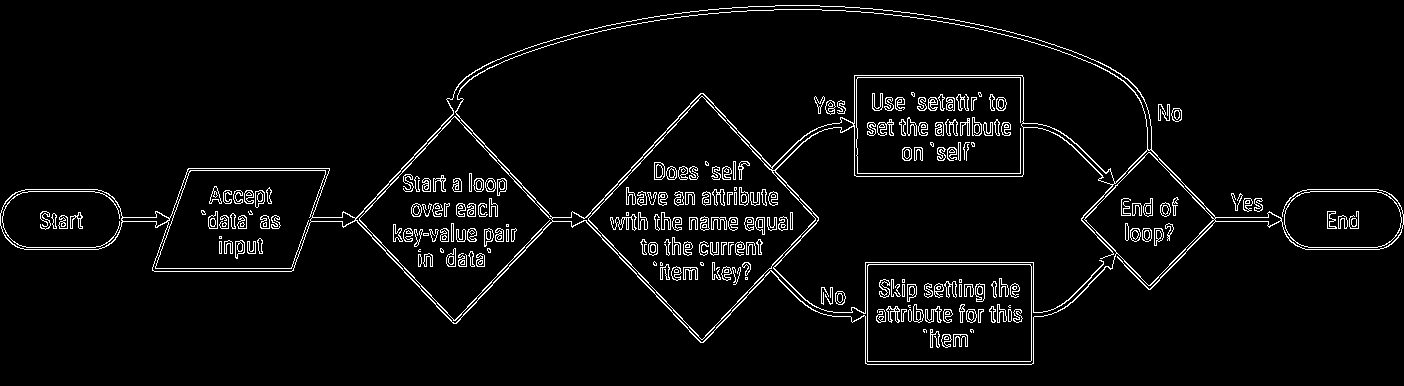}
    \caption{Canny edge map: structural skeleton}
    \label{fig:c3_canny.png}
  \end{subfigure}
\caption{Illustrative example from FlowVQA. The edge map isolates the long loop-back connector, providing a strong geometric cue for the VLM to identify the cyclic topology.}
\label{fig:example_images}
\vspace{-0.4cm}
\end{figure}

Fig.~\ref{fig:example_images} (a) shows a flowchart from FlowVQA~\cite{singh2024flowvqa}\footnote{Original vertical layout adjusted to conserve space.} representing an attribute-initialization loop. Its critical topological feature is the backward edge from the \texttt{End of loop?} node to the \texttt{Start of loop} node. Missing this edge misrepresents the cyclic requirement as a linear one. The corresponding Canny edge map shown in Fig.~\ref{fig:example_images} (b) provides a structural skeleton that highlights this long loop-back connector as a distinct visual cue. Furthermore, Fig.~\ref{fig:mermaid_comparison} illustrates the effect: when the edge map is supplied alongside the original image, the VLM correctly recovers the backward edge (\texttt{G -{}-> C}) that is otherwise missed under the baseline condition.
\vspace{-0.8ex}

\begin{figure}[t]
\centering
\begin{tcolorbox}[colback=white,colframe=gray!60,boxsep=1pt,left=12pt,right=3pt,top=2pt,bottom=2pt] 
\begin{Verbatim}[
    fontsize=\scriptsize,
    commandchars=\\\{\},
    numbers=left,      % Enables line numbering
    numbersep=5pt,     % Distance between numbers and code
    firstnumber=1      % Starts at 1
]
flowchart LR
  A([Start]) --> B[/Accept 'data' as input/]
  B --> C\{Start a loop...\}
  C --> D\{Does 'self' have...\}
  D -- Yes --> E[/Use 'setattr'.../]
  D -- No --> F[/Skip setting.../]
  E --> G\{End of loop?\}
  F --> G
  G -- Yes --> H([End])
  \colorbox{green!25}{G -- No --> C}  \textcolor{gray}{\% recovered by edge map}
\end{Verbatim}
\end{tcolorbox}
\caption{Qwen3-VL-32B generated Mermaid code for the FlowVQA example (Fig.~\ref{fig:example_images} (a)). }

\label{fig:mermaid_comparison}
\vspace{-1em}
\end{figure}

\subsection{\textbf{Contributions}}
The contributions of this article are as follows:

\textbf{1) EdgeFlow: VLM-based flowchart processing with edge-map augmentation.}
We propose \textsf{EdgeFlow}, a training-free framework that augments a VLM with \emph{deterministically extracted} Canny edge maps--acting as structural priors--to improve the topological fidelity of flowchart-to-Mermaid conversion without requiring any annotated diagram data.

\textbf{2) Empirical evaluation with statistical rigor.}
We evaluate \textsf{EdgeFlow} on \textit{IndusReqFlow} (52 industrial flowcharts) at the node, edge, and path levels, using per-flowchart Wilcoxon signed-rank tests and Cliff's Delta effect sizes. The results demonstrate the effectiveness of \textsf{EdgeFlow}. 
We also conducted cross-dataset evaluation on a subset of FlowVQA~\cite{singh2024flowvqa} and the results indicate the need for diverse benchmarks for future RE tool evaluation.

\section{\textbf{Background}}\label{sec:background}

In this study, we use the diagrams broadly to encompass the wide variety of graphical models used in requirements engineering (e.g., UML). Our proposed approach, however, specifically targets flowcharts, a fundamental subset of diagrams characterized by directed graphs that represent sequential control-flow logic. 
\subsection{\textbf{Vision Language Models}}\label{sec:vlm}
Vision Language Models (VLMs) combine an image encoder with an LLM to support tasks that require jointly interpreting visual content and generating or reasoning in natural language. 

Many state-of-the-art VLMs utilize a Vision Transformer (ViT) as their visual backbone. ViT architectures represent an image as a sequence of fixed-size patches, applying Transformer self-attention to model interactions among these patches. While this supports globally contextualized visual features, fixed-size patch tokenization is often too coarse to preserve the high-frequency geometric signals critical for flowchart topology. Specifically, thin connectors and small arrowheads, which encode control-flow direction, frequently become fragmented across patch boundaries. This discretization leads to a loss of salience in the pooled visual features, making the resulting graph models prone to missing or incorrect edges.

\subsection{\textbf{Canny Edge Detection and Structural Priors}}\label{sec:bg_canny}
Edge detection is a classical computer vision operation that highlights locations of strong intensity change, producing an edge map that emphasizes boundaries and thin line structures. The Canny detector is a widely adopted method that (i)~smooths the image, (ii)~computes intensity gradients, (iii)~applies non-maximum suppression to localize edges, and~(iv) uses hysteresis thresholding to retain coherent edge contours while suppressing noise~\cite{canny1986computational}.

As flowcharts encode control-flow primarily through line segments (connectors) and arrowheads, Canny edge maps can isolate these topology-critical features from the surrounding visual content (colors, fills, text). In the computer vision and machine learning literature, such a deterministic, domain-informed signal that is supplied alongside a primary input to bias model inference is called a \emph{structural prior}~\cite{zhang2023adding}. EdgeFlow adapts the same principle to flowchart processing in an RE context. Section~\ref{sec:methodology} describes how our approach exploits this prior by supplying the Canny edge map as a second visual input alongside the original flowchart image.

\section{\textbf{Methodology}}
\label{sec:methodology}
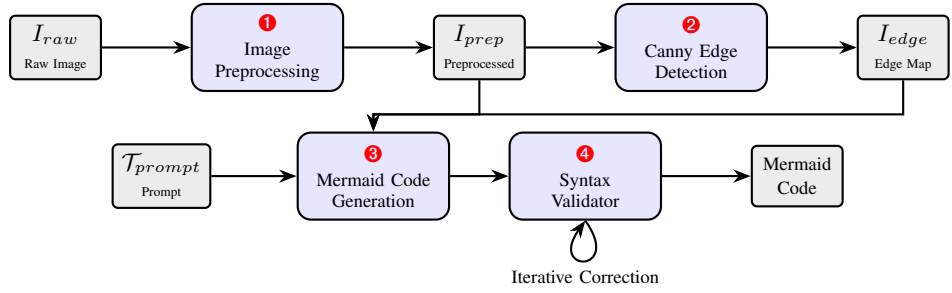
\begin{figure*}[t]
	\centering
	\begin{tikzpicture}[
			font=\small,
			imgnode/.style={draw, thick, fill=gray!15, rounded corners=2pt,
					minimum width=1.2cm, minimum height=0.8cm, align=center},
			stepnode/.style={draw, thick, fill=blue!10, rounded corners=5pt,
					minimum width=2.0cm, minimum height=1.15cm, align=center},
			outnode/.style={draw, thick, fill=green!12, rounded corners=4pt,
					minimum width=1.5cm, minimum height=0.8cm, align=center},
			arr/.style={-Stealth, thick},
		]

		\node[imgnode] (iraw)  at (0,0)   {$I_{raw}$\\{\tiny Raw Image}};
		\node[stepnode] (step1) at (2.8,0)  {\textbf{\StepOne}\\[-1pt]{\scriptsize Image}\\[-2pt]{\scriptsize Preprocessing}};
		\node[imgnode] (iprep) at (5.6,0)  {$I_{prep}$\\{\tiny Preprocessed}};
		\node[stepnode] (step2) at (8.4,0)  {\textbf{\StepTwo}\\[-1pt]{\scriptsize Canny Edge}\\[-2pt]{\scriptsize Detection}};
		\node[imgnode] (iedge) at (11.2,0)  {$I_{edge}$\\{\tiny Edge Map}};

		\node[imgnode]  (prompt)    at (1.4,-1.7)  {$\mathcal{T}_{prompt}$\\{\tiny Prompt}};
		\node[stepnode] (step3)     at (4.2,-1.7)  {\textbf{\StepThree}\\[-1pt]{\scriptsize Mermaid Code}\\[-2pt]{\scriptsize Generation}};
		\node[stepnode]  (step4) at (7.0,-1.7)  {\textbf{\StepFour}\\[-1pt]{\scriptsize Syntax}\\[-2pt]{\scriptsize Validator}};
		\node[imgnode]  (output)    at (9.8,-1.7) {{\scriptsize Mermaid}\\[-1pt]{\scriptsize Code}};

		\draw[arr] (iraw)  -- (step1);
		\draw[arr] (step1) -- (iprep);
		\draw[arr] (iprep) -- (step2);
		\draw[arr] (step2) -- (iedge);
		\draw[arr] (iprep.south) -- ++(0,-0.45) -| (step3.north);
		\draw[arr] (iedge.south) -- ++(0,-0.45) -| (step3.north);
		\draw[arr] (prompt)    -- (step3);
		\draw[arr] (step3)     -- (step4);
		\draw[arr] (step4) -- (output);
		\draw[arr] (step4.south) .. controls +(0.5,-0.7) and +(-0.5,-0.7) .. (step4.south) node[midway, below] {\scriptsize Iterative Correction};

	\end{tikzpicture}
    \setlength{\belowcaptionskip}{-10pt}
	\caption{Overview of our four-step approach:
		(\StepOne)~Image Preprocessing standardizes the raw flowchart image $I_{raw}$ into $I_{prep}$;
		(\StepTwo)~Canny Edge Detection extracts the binary structural edge map $I_{edge}$;
		(\StepThree)~VLM-based Mermaid Code Generation synthesizes the machine-readable representation
		from both $I_{prep}$, $I_{edge}$, and a structured prompt $\mathcal{T}_{prompt}$;
		(\StepFour)~Syntax Validation ensures the generated code conforms to Mermaid specifications.}
	\label{fig:approach}
	\vspace{-0.5em}
\end{figure*}

Figure~\ref{fig:approach} presents an overview of our approach, which consists of four steps: (\StepOne) ~Image Preprocessing, (\StepTwo) ~Deterministic Structural Prior Extraction, (\StepThree) ~Mermaid Code Generation, and (\StepFour) ~Syntax Validation.

\sectopic{Step~1)~ Image Preprocessing}
\label{subsec:preprocessing}

To handle the heterogeneous formats, varying resolutions, and compression artifacts typical of industrial documentation, we apply a standardization pipeline formally defined as $I_{prep} = \mathcal{P}(I_{raw})$. This step normalizes visual features to ensure consistent VLM behavior across the dataset:

\textbf{Alpha Channel Normalization}: Some industrial flowchart images contain transparency information (a fourth ``alpha'' channel beyond the standard RGB color channels). When such images are fed to a VLM, transparent regions are rendered as black by default, creating spurious dark areas that the model may misinterpret as diagram elements. We composite all images onto a white background to eliminate this artifact.

\textbf{Adaptive Rescaling}: We downscale images so that no dimension exceeds 4{,}000 pixels, using high-quality interpolation to preserve fine-grained details such as arrowheads and connector text that are critical for topological analysis, while keeping the input image within the resolution limits of VLMs which are deployed practically for Huawei's test engineers.

\vspace{7.5pt}
\sectopic{Step~2)~ Deterministic Structural Prior Extraction}
\label{subsec:edge_detection}

In this step, we extract high-frequency structural signals $I_{edge}$ from the preprocessed image produced in Step~\StepOne{} using the Canny edge detection algorithm introduced in Section~\ref{sec:bg_canny}. Since Canny detection depends solely on pixel intensity gradients, it ensures that identical visual inputs yield identical structural maps without stochastic inference noise. The process is formally defined as $I_{edge} = \textit{Canny}(I_{prep}, \theta)$, where $\theta$ represents the hysteresis thresholding and aperture hyperparameters. 
While $I_{prep}$ retains semantic content, $I_{edge}$ acts as a high-contrast topological scaffold that isolates connectivity features--such as thin line segments and arrowheads--from the diagram's background metadata.

The hysteresis thresholds and Sobel aperture size are treated as tunable hyperparameters; their selection through empirical optimization is detailed in Section~\ref{sec:E-rq_procedure}.

\sectopic{Step~3)~ Mermaid Code Generation}
\label{subsec:mermaid_generation}

In this stage, we synthesize the machine-readable representation using a composite visual prompting strategy. We construct a multi-image input tuple $\mathcal{X} = \langle I_{prep}, I_{edge}, \mathcal{T}_{prompt} \rangle$, supplying the VLM with both the original semantic context ($I_{prep}$) and the explicit structural scaffold ($I_{edge}$) extracted in Step~\StepTwo{}. We adopt a zero-shot prompting strategy with structured prompts composed of a system role definition and a user task instruction. The prompt templates are as follows:

\begin{tcolorbox}[colback=blue!5!white,colframe=blue!75!black,
        title=System Prompt: Mermaid Code Generation Role Definition,
        fonttitle=\bfseries\footnotesize,fontupper=\scriptsize,
        boxsep=2pt,left=3pt,right=3pt,top=3pt,bottom=3pt,
        before skip=2pt,after skip=2pt]
	\textbf{Role}: You are a professional visual analysis agent specialized in flowchart analysis and Mermaid code generation.\\
	\textbf{Guidelines to Follow}: (i) Observe the overall structure and layout of flowcharts comprehensively; (ii) Identify main node types and their distributions accurately; (iii) Analyze text labels to understand their meaning and function; (iv) Infer complete flow logic and branching structures based on visual evidence; (v) Generate accurate, executable Mermaid flowchart code.
\end{tcolorbox}

\begin{tcolorbox}[colback=green!5!white,colframe=green!75!black,
        title=User Prompt: EdgeFlow (Dual-Image) Mermaid Code Generation,
        fonttitle=\bfseries\footnotesize,fontupper=\scriptsize,
        boxsep=2pt,left=3pt,right=3pt,top=3pt,bottom=3pt,
        before skip=2pt,after skip=6pt]
	\textbf{Task}: Analyze the given flowchart image pair and generate corresponding Mermaid flowchart code.\\[2pt]
	\textbf{Instruction}: Your task involves multiple images--the preprocessed flowchart and its edge-detected version. First, observe the structural edge map to identify connectivity details such as thin lines, long edges, and arrowheads. Second, cross-reference with the preprocessed image to extract node labels and semantic content. Finally, synthesize complete Mermaid code that captures both topological structure and semantic information. Ensure the generated code follows Mermaid syntax conventions and represents the full flow logic including decision branches and loop structures. The Mermaid code must be enclosed in triple backticks with language specifier (i.e., \texttt{\textbackslash\textbackslash\textbackslash`mermaid ... \textbackslash\textbackslash\textbackslash`}).\\[2pt]
	\textbf{Input}: [Preprocessed Flowchart Image] [Canny Edge Map Image]
\end{tcolorbox}

\sectopic{Step~4)~ Syntax Validation}
\label{subsec:syntax_validation}
In the final stage, we ensure the syntactic validity of the generated Mermaid code through an iterative validation-correction cycle. The initial code generated in Step~\StepThree{} first undergoes automated preprocessing to remove non-functional comments and unsupported syntax. We then validate the cleaned code using the \texttt{mermaid-ast} parser~\cite{emily_mermaid_ast_2026}. If parsing succeeds, the syntactically valid Mermaid code is returned as \textsf{EdgeFlow} pipeline's final output. If parsing fails, the parser produces a structured error log containing diagnostic messages and line numbers, which is passed to a code-specialized LLM with instructions to correct syntax errors while preserving the original business logic, node labels, and edge relationships. This iterative loop within Step~\StepFour{} repeats until parsing succeeds or a maximum threshold of 10 iterations is reached--a limit empirically established to ensure timely correction in production environments. 

\section{\textbf{Experimental Setup}}\label{sec:evaluation}
This section presents the experimental setup for evaluating our approach, including the research questions, dataset construction, evaluation metrics, baseline configuration, implementation details, and analysis procedures.

\subsection{\textbf{Research Questions}}
To evaluate the effectiveness and generalizability of \textsf{EdgeFlow}, we investigate the following research questions:

\begin{itemize}
	\item \textbf{RQ1.} Does \textsf{EdgeFlow}, by integrating deterministic structural visual priors (Canny edge maps), improve topological extraction correctness (nodes and edges) in industrial flowcharts compared to vanilla VLMs?
	\item \textbf{RQ2.} To what extent does \textsf{EdgeFlow} enhance path generation for downstream model-based testing (MBT) tasks, as measured by path-level correctness?
	\item \textbf{RQ3.} How does the performance gain of \textsf{EdgeFlow} generalize from noisy industrial datasets to publicly available clean synthetic datasets?
\end{itemize}

\subsection{\textbf{Dataset}} \label{sec:dataset}
We construct two flowchart datasets for comprehensive evaluation across different domains: (1)~\textbf{IndusReqFlow}, a proprietary industrial dataset for in-domain evaluation (RQ1, RQ2), and (2)~\textbf{a subsample of FlowVQA}~\cite{singh2024flowvqa}, a publicly available synthetic dataset for cross-domain generalization assessment (RQ3).

\subsubsection{\textbf{IndusReqFlow}} \label{sec:indusreqflow}To bridge the synthetic-to-real domain shift gap, we curated \textit{IndusReqFlow}, a dataset of \textbf{52} real-world industrial flowcharts sourced from Huawei's system-level requirement documents. The dataset spans multiple domains: optical networks, data communications, and electric vehicle systems. The ground truth annotation process involved two phases. First, test engineers individually authored Mermaid code for each flowchart. Second, they validated these annotations by comparing the Mermaid-rendered output against the original image to ensure node and edge consistency. Disagreements were identified in 8 out of 52 flowcharts (15.4\%), due to non-standard notations or additional labels; 3 cases were then resolved through discussion between the two primary annotators, while the remaining 5 required consultation with a third test engineer for domain expert judgment. Across the 52 flowcharts, \textit{IndusReqFlow} contains an average of 11.15 nodes and 11.73 edges per flowchart. Due to confidentiality and proprietary constraints of Huawei's internal documents, the IndusReqFlow dataset is not publicly available.

\subsubsection{\textbf{FlowVQA (Public Dataset)}}\label{sec:flowvqa}
For cross-domain evaluation, we randomly sampled \textbf{40} synthetic flowcharts from FlowVQA~\footnote{Samples and results:\url{https://github.com/ZhifeiDou/EdgeFlow-RE2026}}, a publicly available flowchart dataset\cite{singh2024flowvqa}. FlowVQA provides ground-truth Mermaid code for each flowchart, eliminating the effort for human annotation. The selection was performed randomly to obtain a representative subset of FlowVQA in terms of node count and diagram layout. 
The sampled FlowVQA dataset averages 21.35 nodes and 23.70 edges per flowchart.

Visual characteristics of \textit{IndusReqFlow} vs. sampled \textit{FlowVQA} reflect the challenges of industrial documentation:

\textbf{Higher visual noise}: Compared to synthetic diagrams, industrial flowcharts exhibit higher background noise\footnote{Background noise ($\sigma$): standard deviation of high-frequency components after Gaussian blur subtraction in flat image regions, measuring scanning sensor noise and compression artifacts. Background color instability: mean chromatic variation in light background regions using CIELAB color space, measuring uneven illumination and color inconsistency.} (industrial: $\sigma=33.79$ vs.\ synthetic: $\sigma=19.86$, ratio $=1.70\times$) and greater background color instability (industrial: mean$=5.38$ vs.\ synthetic: mean$=2.67$, ratio $=2.02\times$), attributable to scanning artifacts, photocopying degradation, paper aging, and varying illumination conditions across acquisition sessions.

\textbf{Structural ambiguity}: Industrial dataset also exhibits greater ambiguity in structural elements, such as overlapping connectors, crossed lines without jump markers, and non-standard notations, which  create interpretation challenges for semantic-reliant models.

\begin{table}[htbp]
    \centering
    \captionsetup{skip=2pt}
    \caption{Comparison of Datasets Used in This Study.}
    \label{tab:dataset-comparison}
    \footnotesize 
    \setlength{\tabcolsep}{3pt}
    \begin{tabularx}{\columnwidth}{@{}l l c c c l p{3.2cm}@{}}
        \toprule
        \textbf{Dataset} & \textbf{Source} & \textbf{Size} & \textbf{Avg N} & \textbf{Avg E} & \textbf{Noise} & \textbf{Noise Metrics} \\
        \midrule
        FlowVQA \cite{singh2024flowvqa} & Public & 40 & 21.35 & 23.70 & Low & $\sigma=19.86, \mu=2.67$ \\
        IndusReqFlow & Indust. & 52 & 11.15 & 11.73 & High & $\sigma=33.79, \mu=5.38$ \\
        \bottomrule
    \end{tabularx}
    \vspace{-0.4cm} 
\end{table}

\subsection{\textbf{Evaluation Metrics}}\label{subsec:eval_metrics}
We evaluate correctness with three metric levels--\textbf{node}, \textbf{edge}, and \textbf{path}--by comparing the generated Mermaid code against the ground-truth, reporting precision ($P$), recall ($R$), and F1-score at each level.

\subsubsection{\textbf{Matching Criteria}}\label{sec:matching_criteria}
All three levels employ exact string matching for label comparison (case-sensitive, without normalization) under a greedy one-to-one constraint. This strict criterion reflects the practical requirements of our collaborating test engineers, as even single-character deviations in industrial flowchart labels (e.g., \texttt{PortA} vs.\ \texttt{PortB}) can denote distinct subsystems or test conditions.

\paragraph{\textbf{Node-Level}}
A predicted node is a $TP$ if and only if its label is identical to a ground-truth node label.

\paragraph{\textbf{Edge-Level}}
Edges are represented as ($source\_label$,\; $target\_label$,\;$edge\_label$) tuples. A $TP$ edge requires all three components to match exactly with a ground truth edge. 

\paragraph{\textbf{Path-Level}}\label{sec:path_metrics}
Path-level evaluation corresponds to  the downstream MBT workflow in which test engineers derive test cases from flowcharts~\cite{briand2002umltesting,utting2010practicalmbt}, as each extracted path corresponds to a candidate test case. A $TP$ path is a predicted path whose node-label sequence is identical to a ground-truth.

\subsubsection{\textbf{Aggregation and Statistical Testing}}\label{sec:stat_testing}
We report results at two aggregation levels, both computing $P$, $R$, and $F1$ separately per metric level (node, edge, and path, respectively).

\paragraph{\textbf{Global Level}}
For each metric level, the $TP$, $FP$, and $FN$ counts are aggregated across all flowchart–batch pairs, and micro-averaged precision, recall, and F1-score are computed from the aggregated results.

\paragraph{\textbf{Per-Flowchart-Level}}

For a given flowchart, the $TP$, $FP$, and $FN$ counts at each metric level are aggregated across all repeated experimental runs. Precision, recall, and F1-score are computed from these per-flowchart aggregates. This yields one score per flowchart per metric level per experimental condition, producing $N$ paired observations that serve as input to the subsequent  statistical tests below ($N=52$ for IndusReqFlow, $N=40$ for FlowVQA).

To assess statistical significance, we apply the Wilcoxon signed-rank test~\cite{wilcoxon1945individual}, a non-parametric paired test requiring no normality assumption. We report one-sided $p$-values ($H_1$: EdgeFlow $>$ Baseline) and complement them with Cliff's Delta ($\delta$)~\cite{cliff1993dominance} as an effect size measure, using the standard cutoffs: $|\delta|{<}0.147$ negligible, ${<}0.33$ small, ${<}0.474$ medium, ${\geq}0.474$ large. We additionally report win/tie/loss counts across individual flowcharts.

\subsection{\textbf{Implementation}}\label{sec:implementation}

The design and reporting of our LLM-based approach follows the guidelines for empirical studies involving LLMs by Baltes et al.~\cite{baltes2025guidelines}. Specifically, we report model identifiers, parameter counts, generation parameters, complete prompt templates (Section~\ref{subsec:mermaid_generation}), and the number of experimental runs.

\paragraph{\textbf{Models and Platform}}
To reflect test engineers' daily workflow environment, all models are accessed through Huawei's enterprise infrastructure platform running on Ascend NPUs. From the models available on this platform, we select two VLMs for Mermaid Code Generation step (\StepThree): \textbf{Qwen3-VL-32B-Instruct} \footnote{We use Qwen3-VL-32B to represent Qwen3-VL-32B-Instruct for brevity.} (dense, 32B parameters) and \textbf{Qwen3.5-35B-A3B} (sparse MoE, 35B total/3B active parameters), representing dense and mixture-of-experts (MoE) architectures respectively. For syntax validation step (\StepFour), we adopt the code-specialized LLM available on platform during experimentation \textbf{Qwen3-Coder-Next} (sparse MoE, 80B total/3B active parameters). All models run at full precision with fixed generation parameters: \texttt{temperature=0.3}, \texttt{top\_p=0.8}, \texttt{max\_tokens=16000}, to reduce output variability.

\paragraph{\textbf{Baseline}}
The baseline condition isolates the contribution of the Canny edge map by providing the VLM with only the preprocessed image $I_{prep}$---without the edge map---while keeping all other variables identical (i.e.,  same model, generation parameters, system prompt, and same syntax validation step). The baseline Mermaid code generation user prompt follows the similar structure as the \textsf{EdgeFlow} prompt (in Section~\ref{subsec:mermaid_generation}), differing in receiving a single image input.

\begin{tcolorbox}[colback=yellow!5!white,colframe=yellow!75!black,
		title=User Prompt: Baseline (Single-Image) Mermaid Code Generation,
		fonttitle=\bfseries\footnotesize,fontupper=\scriptsize,
		boxsep=2pt,left=3pt,right=3pt,top=3pt,bottom=3pt,
		before skip=1pt,after skip=1pt]
	\textbf{Task}: Analyze the given flowchart image and generate corresponding Mermaid flowchart code.\\[2pt]
	\textbf{Instruction}: Your task involves a single flowchart image. Observe the image comprehensively to identify nodes (including their labels) and connectivity details such as edges and arrowheads. Generate complete Mermaid code that captures the full flow logic including decision branches and loop structures. Ensure the generated code follows Mermaid syntax conventions. The Mermaid code must be enclosed in triple backticks with language specifier (i.e., \texttt{\textbackslash\textbackslash\textbackslash`mermaid ... \textbackslash\textbackslash\textbackslash`}).\\[2pt]
	\textbf{Input}: [Preprocessed Flowchart Image]
\end{tcolorbox}

\subsection{\textbf{Analysis Procedure}}\label{sec:E-rq_procedure}
Each research question follows a consistent protocol: five independent experimental runs per condition on the target dataset--where inter-run variability arises from the non-zero sampling temperature (0.3)--with results aggregated and tested for statistical significance as defined in Section~\ref{subsec:eval_metrics}

\begin{table*}[htbp]
	\centering
    \captionsetup{skip=2pt}
	\caption{Canny parameter configurations for the two-stage optimization on IndusReqFlow.}
	\label{tab:canny-complete-configurations}
    \begin{tabular}{ccccll}
		\toprule
		\textbf{Config} & \textbf{Low\textsuperscript{$\dagger$}} & \textbf{High\textsuperscript{$\dagger$}} & \textbf{Aperture} & \textbf{Model} & \multicolumn{1}{c}{\textbf{Rationale}} \\
		\midrule
		\multicolumn{6}{l}{\textit{Stage~1: Threshold selection (aperture fixed at 3)}} \\
		\midrule
		C1 & 30  & 100 & 3 & Both VLMs & Low thresholds retaining connectors degraded by compression artifacts and resolution variance \\
		C2 & 50  & 150 & 3 & Both VLMs & Intermediate thresholds balancing edge retention against compression noise \\
		C3 & 100 & 200 & 3 & Both VLMs & ControlNet default~\cite{zhang2023adding}; reproducible, externally validated starting point \\
		C4 & 100 & 300 & 3 & Both VLMs & Conservative thresholds retaining only prominent edges; tests minimal skeleton sufficiency \\
		\midrule
		\multicolumn{6}{l}{\textit{Stage~2: Aperture selection (thresholds fixed from Stage~1)}} \\
		\midrule
		C6 & 100 & 200 & 5 & Qwen3-VL-32B    & Increased smoothing over Stage~1 default (aperture 3) \\
        C7 & 100 & 200 & 7 & Qwen3-VL-32B    & Maximum Sobel aperture; strongest smoothing level \\
        C8 & 50  & 150 & 5 & Qwen3.5-35B-A3B & Increased smoothing on MoE-optimal thresholds \\
        C9 & 50  & 150 & 7 & Qwen3.5-35B-A3B & Maximum Sobel aperture on MoE-optimal thresholds \\
		\bottomrule
	\end{tabular}\\[2pt]
	{\footnotesize \textsuperscript{$\dagger$}Canny hysteresis threshold lower and upper bounds.}
    \vspace{-0.4cm} 
\end{table*}

\subsubsection{\textbf{Analysis Procedure for RQ1}}\label{sec:rq1_procedure}
RQ1 is addressed in two phases: a) we determine \textsf{EdgeFlow}'s optimal Canny configuration for each VLM; b)  we compare \textsf{EdgeFlow} against the baseline.

\paragraph{\textbf{Canny Parameter Optimization}}
We determine \textsf{EdgeFlow}'s Canny hyperparameters via a two-stage optimization, varying one parameter group at a time while holding all other experimental variables from Implementation (Section~\ref{sec:implementation}) constant (such as VLM architecture, generation parameters, and prompt templates). We evaluate configurations from C1 to C9, with their parameters and rationale listed in Table~\ref{tab:canny-complete-configurations}. Each configuration is evaluated across five independent runs on \textit{IndusReqFlow} (52 flowcharts per run, 260 samples total). For each run, flowchart images are converted to Mermaid code by adopting the \textsf{EdgeFlow} pipeline in Methodology (Section~\ref{sec:methodology}). The resulting Mermaid Code is evaluated against ground-truth using the matching criteria in Section~\ref{sec:matching_criteria}.

\textbf{Stage~1 (Threshold Selection).} With the Sobel aperture fixed at 3, we test four hysteresis threshold low/high pairs on both VLMs (configurations C1 through C4 in Table~\ref{tab:canny-complete-configurations}), maintaining the recommended hysteresis ratios of 2:1 and 3:1. For each VLM, the best configuration is determined according to the global-level micro-averaged node and edge $F1$ defined in Section~\ref{subsec:eval_metrics}; then serves as reference configuration (C5, omitted in all following tables) in next stage.

\textbf{Stage~2 (Aperture Selection).} Using thresholds of \textbf{Stage~1 optimal configuration} for each VLM, we vary the Sobel aperture size to determine the smoothing level that maximizes topological fidelity. For Qwen3-VL-32B, which utilize C3 as threshold reference configuration, we fix low=100/high=200 and test apertures 5 and 7 as C6 and C7. For Qwen3.5-35B-A3B, which adopt C2 as threshold reference configuration, we fix low=50/high=150 and test apertures 5 and 7 as C8 and C9. The configuration with highest global-level micro-averaged node and edge $F1$ across two stages is determined as final optimal configuration and adopted for the baseline comparison in Section~\ref{Baseline Comparison}. 

\paragraph{\textbf{Baseline Comparison}}\label{Baseline Comparison}
For each VLM, we conduct five independent baseline runs defined in Implementation (Section~\ref{sec:implementation}) on \textit{IndusReqFlow}. We then compare the result against the result of EdgeFlow's final optimal configuration from Stage 2, following the aggregation and statistical analysis metrics in Section~\ref{sec:stat_testing}.

\subsubsection{\textbf{Analysis Procedure for RQ2}}\label{sec:rq2_procedure}
RQ2 assesses whether \textsf{EdgeFlow}'s node- and edge-level improvements from RQ1 enable  better support for downstream MBT. In MBT workflows that use flowcharts as the behavioral model, each entry-to-terminal path through the control-flow graph is used to construct test cases~\cite{briand2002umltesting,utting2010practicalmbt}; consequently, path-level correctness reflects the quality of automatically derived test cases. For each VLM, we reuse the Mermaid code produced in RQ1--five independent \textsf{EdgeFlow} runs under the optimal configuration (C3 for Qwen3-VL-32B, C9 for Qwen3.5-35B-A3B) and five independent baseline runs--yielding 260 samples per condition (52 flowcharts $\times$ 5 runs). To extract test case paths, each Mermaid Code is parsed into a JSON structure of nodes and edges, from which an adjacency list is constructed. Entry nodes and terminal nodes are identified, and a depth-first search enumerates all entry-to-terminal paths; for flowcharts containing cycles, each loop is unrolled once, which is consistent with the standard MBT practice of requiring at least one iteration through each loop body for minimal path coverage~\cite{utting2010practicalmbt}. Ground truth paths are extracted from the ground-truth Mermaid Code using the same procedure. Path-level $P$, $R$, and $F1$ are computed following path-level metrics defined in Section~\ref{sec:path_metrics}, with aggregation and statistical testing per Section~\ref{sec:stat_testing}.

\subsubsection{\textbf{Analysis Procedure for RQ3}}\label{sec:rq3_procedure}
To assess whether \textsf{EdgeFlow} generalizes beyond industrial data, we apply the baseline comparison protocol from RQ1 to the FlowVQA subsample (Section~\ref{sec:flowvqa}), evaluating node- and edge-level $P$, $R$, and $F1$. Note that the Canny hyperparameters are \emph{not} re-optimized for FlowVQA --  we reuse the optimal configurations identified on IndusReqFlow (C3 for Qwen3-VL-32B, C9 for Qwen3.5-35B-A3B) to test generalization without dataset-specific tuning. All other variables (VLMs, baseline condition, generation parameters, and evaluation pipeline) remain identical. For both VLMs, we conduct five independent runs (40 flowcharts per run, 200 samples total).

\section{\textbf{Results}}\label{sec:results}
\subsection{\textbf{RQ1 Result: Topological Correctness Improvement}}
\label{sec:rq1}
\paragraph{\textbf{Canny Parameter Optimization Result}}
Prior to the baseline comparison, we conducted a two-stage Canny hyperparameter optimization defined in Section~\ref{sec:E-rq_procedure}. Table~\ref{tab:ablation} summarizes the impact of each Canny configuration on topological extraction performance.
\begin{table*}[htbp]
	\centering
    \captionsetup{skip=2pt}
	\caption{Canny hyperparameter optimization on IndusReqFlow. Bold indicates best metric
    per model.}
	\label{tab:ablation}
	\begin{tabular}{llccccccccc}
		\toprule
		& & & & & \multicolumn{3}{c}{\textbf{Node}} & \multicolumn{3}{c}{\textbf{Edge}} \\
		\cmidrule(lr){6-8} \cmidrule(lr){9-11}
		\textbf{Model} & \textbf{Config} & \textbf{Low\textsuperscript{$\dagger$}} & \textbf{High\textsuperscript{$\dagger$}} & \textbf{Aperture} & \textbf{Precision} & \textbf{Recall} & \textbf{F1} & \textbf{Precision} & \textbf{Recall} & \textbf{F1} \\
		\midrule
		\multirow{6}{*}{Qwen3-VL-32B}
		& C1 & 30  & 100 & 3 & 81.32\% & 70.55\% & 75.55\% & 65.61\% & 58.36\% & 61.77\% \\
		& C2 & 50  & 150 & 3 & 80.86\% & 69.21\% & 74.58\% & 65.44\% & 57.18\% & 61.03\% \\
		& C3 & 100 & 200 & 3 & \textbf{83.51\%} & \textbf{78.93\%} & \textbf{81.16\%} & 66.53\% & \textbf{65.25\%} & \textbf{65.88\%} \\
		& C4 & 100 & 300 & 3 & 83.24\% & 76.55\% & 79.76\% & \textbf{67.44\%} & 64.16\% & 65.76\% \\
		& C6 & 100 & 200 & 5 & 81.23\% & 70.86\% & 75.69\% & 65.37\% & 58.36\% & 61.67\% \\
		& C7 & 100 & 200 & 7 & 83.16\% & 74.59\% & 78.64\% & 67.10\% & 62.20\% & 64.56\% \\
		\midrule
		\multirow{6}{*}{Qwen3.5-35B-A3B}
		& C1 & 30  & 100 & 3 & 60.78\% & 23.03\% & 33.41\% & 41.75\% & 15.84\% & 22.96\% \\
		& C2 & 50  & 150 & 3 & 36.75\% & 39.27\% & 37.97\% & 19.29\% & 27.99\% & 22.84\% \\
		& C3 & 100 & 200 & 3 & 28.49\% & 32.14\% & 30.21\% & 18.80\% & 20.49\% & 19.61\% \\
		& C4 & 100 & 300 & 3 & 24.58\% & 34.83\% & 28.82\% & 18.60\% & 25.64\% & 21.56\% \\
		& C8 & 50  & 150 & 5 & 55.65\% & 47.53\% & 51.27\% & 39.41\% & 29.89\% & 34.00\% \\
		& C9 & 50  & 150 & 7 & \textbf{68.73\%} & \textbf{63.52\%} & \textbf{66.02\%} & \textbf{47.97\%} & \textbf{44.66\%} & \textbf{46.26\%} \\
        
    	\bottomrule
	\end{tabular}\\[2pt]
	{\footnotesize \textsuperscript{$\dagger$}Canny hysteresis threshold lower and upper bounds.}
    \vspace{-0.1cm} 
\end{table*}
For \textbf{Qwen3-VL-32B}, C3 achieved the 
highest node F1 (81.16\%) and edge F1 (65.88\%) across both stages. For \textbf{Qwen3.5-35B-A3B}, C9 substantially outperformed all other configurations with node F1 of 66.02\% and edge F1 of 46.26\%, suggesting that stronger Gaussian smoothing benefits the MoE architecture on \textit{IndusReqFlow}. Across both models, overly low thresholds inflate false-positive edges while overly high thresholds cause edge loss.

\paragraph{\textbf{Baseline Results}}
Table~\ref{tab:performance-comparison} summarizes \textsf{EdgeFlow}'s performance compared to baseline, using the optimal Canny configurations selected above, which are C3 for Qwen3-VL-32B, C9 for Qwen3.5-35B-A3B.

\begin{table*}[htbp]
	\centering
    \captionsetup{skip=2pt}
	\caption{EdgeFlow vs.\ baseline on IndusReqFlow: node- and edge-level performance with statistical significance tests.}
	\label{tab:performance-comparison}
	\begin{tabular}{llcccccc}
		\toprule
		& & \multicolumn{3}{c}{\textbf{Node}} & \multicolumn{3}{c}{\textbf{Edge}} \\
		\cmidrule(lr){3-5} \cmidrule(lr){6-8}
		\textbf{Model} & \textbf{Method} & \textbf{Precision} & \textbf{Recall} & \textbf{F1} & \textbf{Precision} & \textbf{Recall} & \textbf{F1} \\
		\midrule
		\multirow{6}{*}{Qwen3-VL-32B}
		& Baseline (micro-avg) & 67.02\% & 60.83\% & 63.77\% & 51.07\% & 46.98\% & 48.94\% \\
		& EdgeFlow (micro-avg) & 83.51\% & 78.93\% & 81.16\% & 66.53\% & 65.25\% & 65.88\% \\
		\cmidrule(lr){2-8}
		& Baseline (per-flowchart) & 64.54{\scriptsize$\pm$25.01}\% & 63.58{\scriptsize$\pm$23.44}\% & 63.77{\scriptsize$\pm$24.01}\% & 49.31{\scriptsize$\pm$28.01}\% & 49.58{\scriptsize$\pm$28.02}\% & 49.13{\scriptsize$\pm$27.67}\% \\
		& EdgeFlow (per-flowchart) & 81.98{\scriptsize$\pm$13.55}\% & 82.17{\scriptsize$\pm$13.33}\% & 82.03{\scriptsize$\pm$13.33}\% & 65.93{\scriptsize$\pm$20.67}\% & 68.16{\scriptsize$\pm$21.27}\% & 66.87{\scriptsize$\pm$20.62}\% \\
		\cmidrule(lr){2-8}
		& $p$-value & $<$0.001 & $<$0.001 & $<$0.001 & $<$0.001 & $<$0.001 & $<$0.001 \\
		& Cliff's $\delta$ (W/T/L) & 0.71\textsuperscript{L} (43/3/6) & 0.62\textsuperscript{L} (40/4/8) & 0.67\textsuperscript{L} (42/3/7) & 0.67\textsuperscript{L} (43/1/8) & 0.67\textsuperscript{L} (42/3/7) & 0.75\textsuperscript{L} (45/1/6) \\
		\midrule
		\multirow{6}{*}{Qwen3.5-35B-A3B}
		& Baseline (micro-avg) & 60.85\% & 55.79\% & 58.21\% & 37.78\% & 34.85\% & 36.26\% \\
		& EdgeFlow (micro-avg) & 68.73\% & 63.52\% & 66.02\% & 47.97\% & 44.66\% & 46.26\% \\
		\cmidrule(lr){2-8}
		& Baseline (per-flowchart) & 27.13{\scriptsize$\pm$32.54}\% & 27.03{\scriptsize$\pm$32.50}\% & 27.07{\scriptsize$\pm$32.50}\% & 17.42{\scriptsize$\pm$26.14}\% & 17.30{\scriptsize$\pm$25.86}\% & 17.34{\scriptsize$\pm$25.94}\% \\
		& EdgeFlow (per-flowchart) & 66.59{\scriptsize$\pm$22.78}\% & 67.10{\scriptsize$\pm$22.88}\% & 66.79{\scriptsize$\pm$22.75}\% & 49.19{\scriptsize$\pm$30.22}\% & 50.03{\scriptsize$\pm$31.02}\% & 49.53{\scriptsize$\pm$30.47}\% \\
		\cmidrule(lr){2-8}
		& $p$-value & $<$0.001 & $<$0.001 & $<$0.001 & $<$0.001 & $<$0.001 & $<$0.001 \\
		& Cliff's $\delta$ (W/T/L) & 0.60\textsuperscript{L} (40/3/9) & 0.60\textsuperscript{L} (40/3/9) & 0.60\textsuperscript{L} (40/3/9) & 0.60\textsuperscript{L} (38/7/7) & 0.60\textsuperscript{L} (38/7/7) & 0.60\textsuperscript{L} (38/7/7) \\
		\bottomrule
	\end{tabular}

	\vspace{2pt}
    {\footnotesize \textsuperscript{L}Large ($|\delta| \geq 0.474$), \textsuperscript{M}Medium ($\geq 0.33$), \textsuperscript{S}Small ($\geq 0.147$).}
    \vspace{-0.4cm}
\end{table*}

For \textbf{Qwen3-VL-32B}, \textsf{EdgeFlow} achieved a micro-averaged node F1 of 81.16\% and edge F1 of 65.88\%, improving over the baseline by 17.39 percentage points (pp) for node F1 (27.3\% relative) and 16.94\,pp for edge F1 (34.6\% relative). The per-flowchart analysis confirms that these improvements are statistically significant across all metrics ($p < 0.001$, Wilcoxon signed-rank test) with consistently large effect sizes (Cliff's $\delta = 0.67$ for node F1, $\delta = 0.75$ for edge F1). \textsf{EdgeFlow} outperformed the baseline on 42 of 52 flowcharts for node F1 (win/tie/loss: 42/3/7) and 45 for edge F1 (45/1/6), demonstrating consistent improvement and reduced output variability (edge F1 SD: 27.67\,$\to$\,20.62), indicating more stable extraction on heterogeneous industrial inputs. Inspection of the loss cases reveals that they correspond to flowcharts where the baseline already achieves high performance, suggesting that edge-map augmentation provides diminishing returns when the VLM's native perception is already adequate. The larger edge-level relative improvement (+34.6\%) compared to the node level (+27.3\%) underscores that \textsf{EdgeFlow}'s Canny edge priors are effective at recovering connectivity relationships--the core topological challenge in industrial flowcharts.

For \textbf{Qwen3.5-35B-A3B}, \textsf{EdgeFlow} achieved a micro-averaged node F1 of 66.02\% and edge F1 of 46.26\%, improving over the baseline by 7.81\,pp for node F1 (13.4\% relative) and 10.00\,pp for edge F1 (27.6\% relative). Per-flowchart statistical tests confirm significance across all metrics ($p < 0.001$), with consistently large effect sizes (Cliff's $\delta = 0.60$ for both node and edge F1). \textsf{EdgeFlow} improved results on 40 of 52 flowcharts for node F1 (40/3/9) and 38 for edge F1 (38/7/7). Notably, the per-flowchart mean improvement is substantial (+39.72\,pp for node F1 vs.\ +18.26\,pp for Qwen3-VL-32B), reflecting \textsf{EdgeFlow}'s ability to recover outputs on flowcharts where the MoE baseline produced near-zero scores. The MoE model's lower baseline--attributable to limited active parameter capacity (3B active)--provides greater headroom for edge-map augmentation, resulting in larger absolute gains. However, the MoE model shows slightly higher loss counts (9 losses for node F1 vs.\ 7 for the dense model) and lower absolute performance after augmentation (node F1: 66.02\% vs.\ 81.16\%), suggesting that active parameter capacity remains a limiting factor for industrial deployment. 
Despite these architectural differences, both models benefit from \textsf{EdgeFlow}, which we attribute to the high-frequency topological signals extracted by the Canny edge map. By isolating connectivity features as high-contrast patterns, the edge map makes them more visually prominent to the VLM, compensating for the loss of fine-grained geometric detail during ViT patch tokenization.

\vspace{-3pt}
\begin{mdframed}[style=MyFrame]
    \emph{The answer to {\bf RQ1}: \textsf{EdgeFlow} significantly improves topological correctness compared to baseline on IndusReqFlow.}
\end{mdframed}
\vspace{-10pt}

\subsection{\textbf{RQ2: Path Generation for MBT Support}}
\label{sec:rq2}

\begin{table*}[htbp]
	\centering
    \captionsetup{skip=2pt}
	\caption{Path-level test case generation quality 
on IndusReqFlow with statistical significance tests.}
	\label{tab:path-metrics}
	\begin{tabular}{llccc}
		\toprule
		& & \multicolumn{3}{c}{\textbf{Path}} \\
		\cmidrule(lr){3-5}
		\textbf{Model} & \textbf{Method} & \textbf{Precision} & \textbf{Recall} & \textbf{F1} \\
		\midrule
		\multirow{6}{*}{Qwen3-VL-32B}
		& Baseline (micro-avg) & 21.50\% & 15.94\% & 18.31\% \\
		& EdgeFlow (micro-avg) & 33.00\% & 26.45\% & 29.37\% \\
		\cmidrule(lr){2-5}
		& Baseline (per-flowchart) & 17.63{\scriptsize$\pm$26.52}\% & 16.24{\scriptsize$\pm$25.26}\% & 16.47{\scriptsize$\pm$25.10}\% \\
		& EdgeFlow (per-flowchart) & 28.78{\scriptsize$\pm$30.42}\% & 27.35{\scriptsize$\pm$29.05}\% & 27.30{\scriptsize$\pm$28.80}\% \\
		\cmidrule(lr){2-5}
		& $p$-value (eff.\ $n=39$) & $<$0.001 & $<$0.001 & $<$0.001 \\
		& Cliff's $\delta$ (W/T/L) & 0.40\textsuperscript{M} (30/13/9) & 0.50\textsuperscript{L} (32/14/6) & 0.44\textsuperscript{M} (31/13/8) \\
		\midrule
		\multirow{6}{*}{Qwen3.5-35B-A3B}
		& Baseline (micro-avg) & 18.52\% & 24.60\% & 21.13\% \\
		& EdgeFlow (micro-avg) & 26.35\% & 23.90\% & 25.06\% \\
		\cmidrule(lr){2-5}
		& Baseline (per-flowchart) & 14.55{\scriptsize$\pm$24.56}\% & 13.94{\scriptsize$\pm$24.11}\% & 14.09{\scriptsize$\pm$23.90}\% \\
		& EdgeFlow (per-flowchart) & 19.90{\scriptsize$\pm$30.29}\% & 19.28{\scriptsize$\pm$30.12}\% & 19.32{\scriptsize$\pm$29.51}\% \\
		\cmidrule(lr){2-5}
		& $p$-value (eff.\ $n=22$) & 0.005 & 0.038 & 0.005 \\
		& Cliff's $\delta$ (W/T/L) & 0.27\textsuperscript{S} (18/30/4) & 0.17\textsuperscript{S} (15/31/6) & 0.27\textsuperscript{S} (18/30/4) \\
		\bottomrule
	\end{tabular}

	\vspace{2pt}
	{\footnotesize \textsuperscript{L}Large ($|\delta| \geq 0.474$), \textsuperscript{M}Medium ($\geq 0.33$), \textsuperscript{S}Small ($\geq 0.147$).}
    \vspace{-0.4cm} 
\end{table*}

\textbf{Path Coverage Results.}
Using the optimal Canny configurations from RQ1, we evaluated path quality on the full dataset (Table~\ref{tab:path-metrics}). For \textbf{Qwen3-VL-32B}, \textsf{EdgeFlow} improved micro-averaged path F1 by 11.06\, pp. Per-flowchart Wilcoxon signed-rank tests confirm that all three path-level improvements are statistically significant ($p<0.001$), with medium-to-large effect sizes (Cliff's $\delta$: 0.44 for path F1). Notably, path recall achieved a large effect size ($\delta=0.50$), indicating that \textsf{EdgeFlow} recovers a substantially higher proportion of ground-truth execution paths.  Overall, \textsf{EdgeFlow} outperformed the baseline on 31 of 52 flowcharts for path F1.

For \textbf{Qwen3.5-35B-A3B}, \textsf{EdgeFlow} improved micro-averaged precision by 7.83 pp and F1-score from 21.13\% to 25.06\%, though micro-averaged recall decreased by 0.70\,pp--indicating that edge-map augmentation does not uniformly improve path coverage for the MoE model. Per-flowchart Wilcoxon tests confirm significant improvements for precision and F1 ($p=0.005$) and recall ($p=0.038$), all with small effect sizes (Cliff's $\delta$: 0.17--0.27). \textsf{EdgeFlow} outperformed the baseline on 15--18 flowcharts versus 4--6 losses, though 30--31 ties indicate that a substantial proportion of flowcharts showed equivalent path quality under both conditions. These ties likely reflect identical outputs or equivalently zero path matches under both conditions. We note that the absolute path F1 remains moderate (29.37\% for Qwen3-VL-32B), reflecting a fundamental \emph{metric cascade}: under exact-match path evaluation, a single incorrect node label or missing edge invalidates an entire path. Given that even the best node F1 is 81\% and edge F1 is 66\%, the combinatorial propagation of element-level errors to path level is expected.

\begin{mdframed}[style=MyFrame]
    \emph{The answer to {\bf RQ2}: \textsf{EdgeFlow}'s node- and edge-level 
    improvements translate into statistically significant path-level correctness gains for downstream MBT tasks. }
\end{mdframed}
\vspace*{-.75em}

\subsection{\textbf{RQ3: Industrial vs.\ Synthetic Generalization}}\label{sec:rq3}

\begin{table*}[htbp]
	\centering
    \captionsetup{skip=2pt}
	\caption{Cross-dataset evaluation on FlowVQA: 
node- and edge-level performance with statistical 
significance tests.}
	\label{tab:cross-dataset}
	\begin{tabular}{llcccccc}
		\toprule
		& & \multicolumn{3}{c}{\textbf{Node}} & \multicolumn{3}{c}{\textbf{Edge}} \\
		\cmidrule(lr){3-5} \cmidrule(lr){6-8}
		\textbf{Model} & \textbf{Method} & \textbf{Precision} & \textbf{Recall} & \textbf{F1} & \textbf{Precision} & \textbf{Recall} & \textbf{F1} \\
		\midrule
		\multirow{6}{*}{Qwen3-VL-32B}
		& Baseline (micro-avg) & 91.66\% & 91.64\% & 91.65\% & 88.39\% & 89.47\% & 88.93\% \\
		& EdgeFlow (micro-avg) & 93.02\% & 93.02\% & 93.02\% & 89.87\% & 90.90\% & 90.38\% \\
		\cmidrule(lr){2-8}
		& Baseline (per-flowchart) & 91.19{\scriptsize$\pm$23.17}\% & 91.14{\scriptsize$\pm$23.14}\% & 91.16{\scriptsize$\pm$23.15}\% & 88.28{\scriptsize$\pm$25.95}\% & 88.88{\scriptsize$\pm$25.98}\% & 88.54{\scriptsize$\pm$25.90}\% \\
		& EdgeFlow (per-flowchart) & 92.73{\scriptsize$\pm$18.21}\% & 92.73{\scriptsize$\pm$18.21}\% & 92.73{\scriptsize$\pm$18.21}\% & 90.05{\scriptsize$\pm$21.33}\% & 90.63{\scriptsize$\pm$21.30}\% & 90.31{\scriptsize$\pm$21.25}\% \\
		\cmidrule(lr){2-8}
		& $p$-value & 0.069\textsuperscript{ns} & 0.037* & 0.065\textsuperscript{ns} & 0.029* & 0.016* & 0.017* \\
		& Cliff's $\delta$ (W/T/L) & 0.15\textsuperscript{S} (8/30/2) & 0.18\textsuperscript{S} (8/31/1) & 0.15\textsuperscript{S} (8/30/2) & 0.20\textsuperscript{S} (11/26/3) & 0.23\textsuperscript{S} (11/27/2) & 0.23\textsuperscript{S} (12/25/3) \\
		\midrule
		\multirow{6}{*}{Qwen3.5-35B-A3B}
		& Baseline (micro-avg) & 93.39\% & 93.30\% & 93.34\% & 90.62\% & 91.85\% & 91.23\% \\
		& EdgeFlow (micro-avg) & 93.57\% & 93.49\% & 93.53\% & 90.52\% & 91.65\% & 91.08\% \\
		\cmidrule(lr){2-8}
		& Baseline (per-flowchart) & 93.07{\scriptsize$\pm$21.11}\% & 92.98{\scriptsize$\pm$21.07}\% & 93.02{\scriptsize$\pm$21.09}\% & 90.63{\scriptsize$\pm$22.76}\% & 91.36{\scriptsize$\pm$22.72}\% & 90.96{\scriptsize$\pm$22.67}\% \\
		& EdgeFlow (per-flowchart) & 93.11{\scriptsize$\pm$20.05}\% & 93.02{\scriptsize$\pm$20.01}\% & 93.07{\scriptsize$\pm$20.02}\% & 90.37{\scriptsize$\pm$23.20}\% & 91.02{\scriptsize$\pm$23.15}\% & 90.65{\scriptsize$\pm$23.10}\% \\
		\cmidrule(lr){2-8}
		& $p$-value & 0.333\textsuperscript{ns} & 0.333\textsuperscript{ns} & 0.333\textsuperscript{ns} & 0.216\textsuperscript{ns} & 0.200\textsuperscript{ns} & 0.208\textsuperscript{ns} \\
		& Cliff's $\delta$ (W/T/L) & 0.05\textsuperscript{N} (6/30/4) & 0.05\textsuperscript{N} (6/30/4) & 0.05\textsuperscript{N} (6/30/4) & 0.10\textsuperscript{N} (10/24/6) & 0.10\textsuperscript{N} (8/28/4) & 0.10\textsuperscript{N} (10/24/6) \\
		\bottomrule
	\end{tabular}

	\vspace{2pt}
	{\footnotesize \textsuperscript{ns}Not significant ($p \geq 0.05$). *Significant ($p < 0.05$). \textsuperscript{S}Small ($|\delta| \geq 0.147$), \textsuperscript{N}Negligible ($|\delta| < 0.147$). }
    \vspace{-0.4cm} 
\end{table*}

Table~\ref{tab:cross-dataset} reports \textsf{EdgeFlow}'s performance on our subset of FlowVQA\cite{singh2024flowvqa} using the same protocol as RQ1. On FlowVQA, improvements are small to negligible for both VLMs. \textbf{Qwen3-VL-32B} shows statistically significant but small edge-level gains (edge F1: $+$1.45\,pp, $p=0.017$, $\delta=0.23$), while node F1 does not reach significance ($p=0.065$). \textbf{Qwen3.5-35B-A3B} shows no significant difference on any metric (all $p \geq 0.200$, $\delta \leq 0.10$), with 60--75\% of flowcharts producing equivalent results under both conditions. The baseline already exceeds 91\% node F1 on FlowVQA, leaving minimal room for augmentation. Compared  with the large improvements on IndusReqFlow ($\delta \geq 0.62$ for Qwen3-VL-32B). Thus, \textsf{EdgeFlow}'s structural priors are valuable when visual noise and degradation impair VLM perception of connectivity features. 

\begin{samepage}
	\begin{mdframed}[style=MyFrame]
		\emph{The answer to {\bf RQ3}: \textsf{EdgeFlow}'s improvements are substantially weaker on clean synthetic data. The contrasting evaluation results on industrial and synthetic data  highlight the need for diverse benchmarks including real-world industrial data for the comprehensive evaluation of VLM-based RE tools.}
	\end{mdframed}
\end{samepage}
\vspace*{-.75em}

\subsection{\textbf{Implications for RE Practice}}\label{sec:implications}
Three observations from the results above carry directly into industrial deployment. (i)~\emph{Topological errors, not label errors, dominate.} Node F1 is consistently higher than edge F1 across both VLMs (e.g., 63.77\% vs.\ 48.94\% at the Qwen3-VL-32B baseline), confirming that VLMs read text well but misperceive connectivity. Investment in mitigations should therefore target geometric perception. (ii)~\emph{The benefit is largest where it matters most.} \textsf{EdgeFlow}'s gains scale with input degradation: large on noisy industrial scans, small on clean synthetic data. Teams whose requirements documents are predominantly scanned, photocopied, or re-rendered should expect the strongest improvements. (iii)~\emph{Path-level gains translate into MBT support.} The 11-pp path-F1 improvement is statistically significant and consistent across 31 of 52 flowcharts, meaning more entry-to-terminal sequences arrive correct enough to seed candidate test cases without manual repair---a tangible reduction in test-engineer effort given documents that contain dozens of charts under frequent revision. Together, these observations argue that deterministic, training-free image augmentation is a pragmatic first step before considering supervised approaches, and that VLM-based RE tooling should be benchmarked on industrial data rather than synthetic surrogates alone.

\section{\textbf{Related Work}}\label{sec:related}

\subsection{\textbf{Requirements Modeling}}

Visual notations are fundamental to RE, serving as primary media for design and communication~\cite{moody2009physics}. Despite this, RE research has focused heavily on \emph{text-to-model generation}, successfully extracting domain models~\cite{arora2016domain}, UML models~\cite{debari2024uml,eisenreich2025llm,ferrari2024modelgen} from textual requirements. Recovery of structured models from \emph{visual} specifications remains less explored~\cite{deka2025flowchart2mermaid,deka2026bpmn}, as static images effectively strip diagrams of their underlying topological logic--connectivity, branching, and edge direction~\cite{baltes2014sketches}. EdgeFlow addresses this gap by recovering path-enumerable, machine-readable representations from visual flowchart artifacts.

\subsection{\textbf{Flowchart and Diagram Processing}}
\label{sec:existing-approaches}

Flowchart benchmarks~\cite{tannert2023flowchartqa,singh2024flowvqa} reveal significant VLM limitations, notably ``topological blindness''~\cite{pan2024flowlearn} and directional biases favoring standard reading conventions over actual connectivity~\cite{singh2024flowvqa}. Evaluations expose critical barriers to industrial adoption, particularly a brittleness to domain shift where models struggle with noise, compression artifacts, and non-standard notations typical of industrial requirements~\cite{pan2024flowlearn,tannert2023flowchartqa}. Furthermore, they reveal a prevalence of layout--text discrepancy errors, where models hallucinate connectivity by over-relying on label proximity rather than following actual geometric signals like lines and arrowheads~\cite{pan2024flowlearn,singh2024flowvqa}.

Existing mitigations utilize multi-stage pipelines combining detection, OCR, and structural parsing. These systems generate descriptive narratives~\cite{arbaz2024genflowchart}, encode arrow directions into prompts~\cite{omasa2025arrow}, or output Mermaid diagrams~\cite{deka2025flowchart2mermaid} or BPMN representations~\cite{deka2026bpmn}. However, these approaches introduce overhead and error propagation through reliance on detectors and specialized training. In contrast, EdgeFlow adopts deterministic Canny edge detection as a training-free structural prior. Inspired by ControlNet's use of edge-map scaffolds for image generation~\cite{zhang2023adding}, we adapt structural input augmentation for lightweight, deployment-ready flowchart understanding in industrial RE settings.

\section{\textbf{Threats to Validity}}\label{sec:threats}
\subsection{\textbf{Internal Validity}}
\textbf{Parameter selection bias.}
Canny hyperparameters were optimized and evaluated on the same IndusReqFlow dataset (52 flowcharts), as no held-out validation split was employed. This introduces a potential risk of configuration overfitting: the selected parameters (C3 for Qwen3-VL-32B; C9 for Qwen3.5-35B-A3B) may partially reflect dataset-specific characteristics rather than universally optimal settings. We reduce this concern by evaluating only a small, interpretable parameter space (two hysteresis thresholds and one aperture size). The selected configurations are also interpretable: C3 (100/200, aperture 3) corresponds to the ControlNet default~\cite{zhang2023adding}, while C9 (50/150,
aperture 7) applies stronger smoothing for noisier MoE outputs. We further
reuse these configurations on FlowVQA without re-optimization to examine
out-of-domain behavior. Nevertheless, we acknowledge this as a methodological limitation. Future work should validate Canny parameter robustness through cross-validation or evaluation on additional industrial datasets from different organizations.

\vspace{0pt}
\textbf{VLM non-determinism.}
VLM outputs vary across runs due to the non-zero \texttt{temperature}. \textit{Mitigation:} We use a low temperature and fixed \texttt{top\_p=0.8} to reduce output variability, and run each experiment across 5 independent runs to quantify robustness. Results are aggregated via micro-averaging for global metrics and via per-flowchart Wilcoxon signed-rank tests for statistical inference. Model versions and experiment timestamps are recorded to support reproducibility.

\vspace{0pt}
\textbf{Sample size.}
IndusReqFlow comprises 52 industrial flowcharts, which limits statistical power. \textit{Mitigation:} We apply the Wilcoxon signed-rank test on 52 per-flowchart paired observations complemented by Cliff's Delta effect sizes, both of which are appropriate for modest sample sizes. We further report win/tie/loss counts for per-flowchart transparency. For RQ3, we complement our evaluation with 40 synthetic flowcharts from FlowVQA to strengthen cross-domain analysis. We acknowledge that observed trends should be interpreted with this sample size constraint.

\vspace{0pt}
\textbf{Model selection.}
All experiments use models from the Qwen family, constrained by the requirements of the enterprise deployment platform (Section~\ref{sec:implementation}). While EdgeFlow has not yet been tested on other VLM families, we mitigate this by evaluating two architecturally distinct models: a dense model (Qwen3-VL-32B) and a sparse MoE model (Qwen3.5-35B-A3B)--to assess robustness across architectural paradigms.

\subsection{\textbf{External Validity}}

\textbf{Dataset representativeness.}
IndusReqFlow covers three distinct industrial domains (optical networks, data communications, and electric vehicles) but may not generalize to all domains. \textit{Mitigation:} RQ3 complements the industrial evaluation with a public synthetic benchmark to assess cross-domain behavior.

\textbf{Diagram types.}
We evaluate EdgeFlow exclusively on flowcharts. Whether EdgeFlow's deterministic structural prior transfers to other behavioral diagrams--such as sequence diagrams, state machines, or BPMN--remains an open question, as those diagram types rely on different visual conventions for representing connectivity.

\textbf{Choice of visual prior.}
Canny edge detection is a classical, well-understood algorithm, but it is not the only possible structural prior. Generic learned contour detectors could capture different visual cues. We selected Canny for its determinism, zero training cost, and reproducibility; a systematic comparison of alternative priors is left for future work.
\vspace{-4pt}
\subsection{\textbf{Construct Validity}}

\textbf{Topological metrics vs.\ downstream utility.}
Node and edge F1 capture structural correctness, but do not directly measure usefulness for downstream RE tasks. A graph with correct nodes and edges, but incorrect control-flow ordering may score well yet produce unusable test cases. \textit{Mitigation:} We address this gap in RQ2 by evaluating path-level correctness, which directly models the MBT workflow of enumerating paths as candidate test cases.

\textbf{Exact-match strictness.}
Our evaluation uses case-sensitive exact string matching with no normalization (Section~\ref{sec:matching_criteria}). A VLM that generates a semantically correct, but superficially different label (e.g., extra whitespace, minor capitalization differences) would be penalized. \textit{Mitigation:} This strict criterion is deliberately chosen to reflect the industrial requirement of our collaborating test engineers, where even minor label deviations can reference distinct subsystems.

\section{\textbf{Practical Considerations}}\label{sec:practical}
Three deployment-oriented questions tend to appear in industrial adoption discussions. \emph{Latency.} The time for Canny extraction  is negligible compared with the multi-second VLM call. \emph{Training-free vs.\ fine-tuning.} A fine-tuned VLM could in principle outperform \textsf{EdgeFlow} but requires annotated flowchart data, rarely available in proprietary RE settings; \textsf{EdgeFlow} is therefore a strong low-cost baseline that future supervised approaches must clear. \emph{Beyond MBT.} The recovered Mermaid representation also supports other RE activities---traceability between text and visual specifications, change impact analysis on flowchart revisions, and consistency checking across document versions---tasks that currently rely on manual inspection in our partner teams.

\section{Conclusion and Future work}\label{sec:conclusion}

This paper proposes \textsf{EdgeFlow}, a training-free framework that augments VLMs with deterministically extracted Canny edge maps to improve topological correctness in flowchart-to-Mermaid conversion. We  evaluated \textsf{EdgeFlow} on the industrial dataset \textit{IndusReqFlow} and the synthetic dataset \textit{FlowVQA};  the results show statistically significant improvements across the node, edge, and path levels for industrial cases exhibiting  visual noise and degradation issues.

For the dense Qwen3-VL-32B, \textsf{EdgeFlow} improves node F1 by 17.39\,pp and edge F1 by 16.94\,pp ($p < 0.001$, large effect sizes), with consistent gains on 42--45 of 52 flowcharts. At the path level, \textsf{EdgeFlow} improves path F1 from 18.31\% to 29.37\%, translating structural gains into measurably better candidate test cases for MBT. On clean synthetic data
(FlowVQA), improvements are small to negligible, confirming that \textsf{EdgeFlow}'s structural priors are most valuable when visual noise and degradation impair VLM perception. Key insights: (1)~VLMs exhibit systematic topological errors on industrial flowcharts--misperceiving connectivity rather than misidentifying labels--and deterministic edge-map augmentation effectively compensates for this limitation;
(2)~Training-free, deterministic approaches are immediately deployable in industrial settings. Moreover, this finding suggests the necessity of integrating real-world industrial data for the accurate evaluation of VLM-based RE tools.

For future work, we will generalize the structural-prior idea to other behavioral  diagrams; note that state machines and sequence diagrams encode connectivity through different visual conventions, so a learned, notation-aware contour detector may be required. Our experiences shows that a community-curated, multi-organization benchmark would change how VLM-based RE tools are evaluated; to create benchmarks that would let the community fairly compare future tools on industrial flowcharts, we are extending \textit{IndusReqFlow} along these lines.

{

\fontsize{6.6pt}{7.4pt}\selectfont
\bibliographystyle{IEEEtran}
\bibliography{references}
}

\end{document}